\documentclass[twocolumn,prl,showpacs]{revtex4-1}
\usepackage{amsfonts,amsthm}
\usepackage{amssymb,amsmath}

\begin{document}
\title{A reference for the gravitational Hamiltonian boundary term}


\author{Chiang-Mei Chen$^{1}$}\email{cmchen@phy.ncu.edu.tw}
\author{Jian-Liang Liu$^{1}$}\email{liujl@phy.ncu.edu.tw}
\author{James M. Nester$^{1,2,3}$}\email{nester@phy.ncu.edu.tw}
\author{Gang Sun$^{1}$}\email{liquideal@gmail.com}

\address{$^1$Department of Physics \& Center for Mathematics and Theoretical Physics, National Central University, Chungli 320 Taiwan}
\address{$^2$Graduate Institute of Astronomy, National Central University, Chungli 320 Taiwan}
\address{$^3$Institute of Physics, Academia Sinica, Taipei 11529, Taiwan}




\begin{abstract}
The Hamiltonian for physical systems and dynamic geometry generates the evolution of a spatial region along a vector field. It includes a boundary term which not only determines the value of the Hamiltonian, but also, via the boundary term in the variation of the Hamiltonian, the boundary conditions. The value of the Hamiltonian comes from its boundary term; it gives the quasi-local quantities: energy-momentum and angular-momentum/center-of-mass. This boundary term depends not only on the dynamical variables but also on their reference values; these reference values determine the ground state---the state having vanishing quasi-local quantities. Here our concern is how to select on the two-boundary the reference values.  To determine the ``best matched'' reference metric and connection values for our preferred boundary term for Einstein's general relativity, we propose on the boundary two-surface (i) \emph{four dimensional} isometric matching, and (ii) extremizing the value of the energy.
\end{abstract}

\maketitle



Our objective is to find a good reference for the Hamiltonian boundary term.  An important application is to the (quasi-)localization of energy.  It is appropriate to first briefly review some aspects of this topic, especially since energy plays a major role in our strategy.

Energy-momentum is the source of gravity (not just for Einstein's general relativity (GR) but for quite general geometric gravity theories). Gravitating systems can exchange energy-momentum with gravity. This interaction happens \emph{locally}, and energy-momentum is conserved, nevertheless there is no well defined local energy-momentum density for gravity itself. This inescapable conclusion (which may seem somewhat ironic---especially since gravity uniquely detects the local density of energy-momentum due to all other physical sources) was actually established already by Noether in the same paper in which she proved her two famous theorems regarding the role of symmetry in dynamical systems~\cite{Noether}.

This key feature---which can be understood physically as following from the equivalence principle (for a discussion, see~\cite{MTW73}, Section 20.4)---explains why
standard approaches aimed at identifying an energy-momentum density for gravitating systems always led to various non-covariant, reference frame dependent, energy-momentum complexes (generally referred to as \emph{pseudotensors}).  There are two types of ambiguity. First, there was no unique expression, but rather many, including the well known ones found by various investigators~\cite{pseudo},
so which expression should  be used?
And second---since all of these expressions are inherently reference frame dependent---for a chosen expression which reference frame should be used to give the proper physical energy-momentum localization?

The more modern idea is \emph{quasi-local}, i.e., energy-momentum should be associated not with a local density but rather with a closed 2-surface  (for a comprehensive review of this topic see~\cite{Sza09}).

One particular approach to quasi-local energy-momentum is via the Hamiltonian (the generator of time evolution).
It has been shown that the Hamiltonian approach actually includes all the classical pseudotensors as special cases, while taming their inherent problems, by providing clear physical/geometric meaning to the two aforementioned ambiguities~\cite{Chang:1998wj}.


Our research group has developed a covariant Hamiltonian formalism that is applicable to a large class of geometric
gravity theories~\cite{Nester:1991yd,Chen:1994qg,Chen:1998aw,Chang:1998wj,Chen:2000xw,Chen:2005hwa,Nester08}.
For such theories the Hamiltonian 3-form ${\cal H}(N)$---the generator of the evolution of a spatial region along the space-time displacement vector field $N$---is also a conserved Noether current:
\begin{equation}
d{\cal H}(N) \propto \hbox{field eqns} \simeq 0\,.
\end{equation}
It has the general form
\begin{equation}
{\cal H}(N) = N^\mu {\cal H}_\mu + d {\cal B}(N)\,,
\end{equation}
where the 3-form $N^\mu{\cal H}_\mu$---which generates the evolution equations---is, as a consequence of local diffeomorphism invariance, itself proportional to certain field equations (initial value constraints) and thus vanishes ``on shell''. Consequently
the \emph{value} of the Hamiltonian associated with a spatial region $\Sigma$ is determined by the total differential (boundary) term:
\begin{equation}
E(N,\Sigma) := \int_\Sigma{\cal H}(N) = \oint_{\partial\Sigma}{\cal B}(N)\,.
\end{equation}
Since it depends only on the field values on the boundary $S=\partial\Sigma$, this value is \emph{quasi-local}.  With suitable choices of the vector field, it can determine values for  the quasi-local energy-momentum and angular momentum/center-of-mass.

Note that the boundary 2-form ${\cal B}(N)$ can be modified in any way without destroying the conservation property.  (This is a particular case of the usual Noether conserved current ambiguity.)  With this freedom one can arrange for almost any conserved quasi-local values.
Fortunately the Hamiltonian's dynamical role tames that freedom.


One must give consideration to the boundary term in the {variation of the Hamiltonian} (see~\cite{Lan49,Regge:1974zd,KT79}).
Requiring it to vanish yields the \emph{boundary conditions}.
The Hamiltonian is {functionally differentiable} only on the phase space
of fields satisfying these boundary conditions.
Modifying the boundary term changes the boundary conditions.

Some time ago it was found that each of the ``superpotentials'' associated with the classical pseudotensors can serve as the Hamiltonian boundary term for GR.   Thus each pseudotensor corresponds to a Hamiltonian  which evolves the dynamical variables with certain ``built in'' boundary conditions~\cite{Chang:1998wj}.  The differing boundary conditions physically accounts for their differing energy-momentum values.  A similar remark can be made for many of the more modern quasi-local proposals.
Fixing the boundary conditions resolves the first type of ambiguity mentioned above.

Looking more closely into the Hamiltonian boundary term, one must, in general, also introduce into it certain \emph{reference values} which represent the ground state of the field---the ``vacuum'' (or background field) values.
For any quantity $\alpha$ we let $\bar\alpha$ be the reference value.  Our boundary expression will contain terms of the form $\Delta\alpha := \alpha - \bar\alpha$.  Here our concern is how to best select these reference values for GR.


For GR two \emph{covariant-symplectic} boundary terms~\cite{Chen:1994qg} had been identified; one (which was also found at about the same time by Katz, Bi{\v c}\'ak and Lynden-Bell~\cite{KBLB} via a Noether argument using a global reference)
is our preferred choice:
\begin{eqnarray}
{\cal B}(N) = \frac{1}{2\kappa} (\Delta\Gamma^{\alpha}{}_{\beta} \wedge i_N \eta_{\alpha}{}^{\beta} + \bar D_{\beta} N^\alpha \Delta\eta_{\alpha}{}^\beta)\,, \label{B}
\end{eqnarray}
where $\Gamma^\alpha{}_\beta$ is the connection one-form,
$\eta^{\alpha\beta\dots} := * (\vartheta^\alpha \wedge \vartheta^\beta\wedge \cdots)$,  $i_N$
denotes the interior product (or contraction) with the vector field $N$, and $\kappa = 8 \pi G/c^4$.
This choice corresponds to fixing  the orthonormal coframe $\vartheta^\mu$ (equivalently the metric) on the boundary
[this follows since the total differential term in the variation of the Hamiltonian 3-form is
$ di_N (\Delta\Gamma^\alpha{}_\beta \wedge \delta\eta_{\alpha}{}^\beta)$].
At spatial infinity (\ref{B}) gives appropriate expressions for the
energy, momentum, angular-momentum, and center-of-mass~\cite{MTW73,Regge:1974zd,amcom}.
(This is not so special, a large class of other expressions can do this also.)
%
The special virtues of the above expression include
{(i)} at {null infinity} it directly gives the Bondi-Trautman energy and the Bondi energy flux~\cite{Chen:2005hwa},
{(ii)} it is ``covariant'',
{(iii)} it has a positive energy property,
{(iv)} for small spheres it gives a positive multiple of the Bel-Robinson tensor,
{(v)} it yields the first law of thermodynamics for black holes~\cite{Chen:1998aw},
{(vi)}  for spherically symmetric solutions it has the {hoop} property~\cite{O'Murchadha:2009kc}.


For all other fields it is {appropriate} to choose  {vanishing reference values} as the reference ground state---the vacuum.
But for geometric gravity the standard ground state is Minkowski geometry which has a {non-vanishing} metric, so a non-trivial reference is {essential}.  Minkowski geometry is our chosen reference, but we still need to specify exactly which Minkowski geometry.

To explicitly construct a reference, choose, in a neighborhood of the desired spacelike boundary 2-surface $S$, four smooth functions $y^i,\ i=0, 1, 2, 3$  with $dy^0 \wedge dy^1 \wedge dy^2 \wedge dy^3 \ne 0$; these {quasi-Minkowski} coordinates define a Minkowski reference by
\begin{equation}
\bar g = -(dy^0)^2 + (dy^1)^2 + (dy^2)^2 + (dy^3)^2.
\end{equation}
Geometrically, this is equivalent to finding a diffeomorphism embedding a neighborhood of the 2-surface into Minkowski space. The associated reference connection is the pullback of the flat Minkowski connection:
\begin{equation}
\bar\Gamma^\alpha{}_\beta = x^\alpha{}_i (\bar\Gamma^i_{~j} y^j{}_\beta + dy^i{}_\beta) = x^\alpha{}_i dy^i{}_\beta.
\end{equation}
Here $x^\alpha{}_i$ is the inverse of $y^i{}_\alpha$, where $dy^i = y^i{}_\alpha dx^\alpha$.

A Killing field of the reference has the infinitesimal Poincar\'e transformation form
$N^k = \alpha^k + \lambda^k{}_l y^l$, where the translation parameters $\alpha^k$ and the boost-rotation parameters $\lambda_{kl} = \lambda_{[kl]}$ are constants.
For any chosen reference and reference Killing field the 2-surface integral  of the Hamiltonian boundary term gives
\begin{equation}
E(N,S) = \oint_S {\mathcal B}(N) = - \alpha^k p_k(S) + \frac12 \lambda_{kl} J^{kl}(S)\,, \label{E(N,S)}
\end{equation}
which yields both a quasi-local {energy-momentum} and a quasi-local {angular momentum/center-of-mass}.
As long as the reference approaches at an appropriate rate the flat Minkowski space at spatial infinity the integrals $p_k(S),\ J^{kl}(S)$ in the spatial asymptotic limit will agree with accepted expressions for these quantities~\cite{MTW73,Regge:1974zd,amcom}.


For energy-momentum one takes $N$ to be a translational Killing field of the Minkowski reference.  Then the second term in our quasi-local boundary expression~(\ref{B}) vanishes~\cite{remark}.
With $N^k = \alpha^k = $ constant our quasi-local expression now takes the form
\begin{equation}
{\mathcal B}(N) = \alpha^k x^\mu{}_k (\Gamma^\alpha{}_\beta - x^\alpha{}_j \, dy^j{}_\beta) \wedge \eta_{\mu\alpha}{}^\beta\,. \label{B2}
\end{equation}

%

To explicitly determine the specific values of the quasi-local quantities one needs some good way to choose the reference.  Minkowski spacetime is the natural choice, especially for asymptotically flat spacetimes~\cite{otherrefs}.  However, as noted above, almost any four functions will determine some Minkowski reference. With such freedom one can still get almost any value for the quasi-local quantities.  This freedom is the quasi-local version of the second type of ambiguity mentioned in the introduction.

Recently we proposed a program~\cite{ae100} to fix the ``best'' choice of reference.  It has two parts: 4D isometric matching
and optimization of a certain quantity.  Here we present it in more detail---along with a promising alternative optimization.
We have already found that our new procedure works well for an important special case: a certain class of axisymmetric spacetimes~\cite{cjp13}.


We first recall a somewhat simpler but still quite important procedure that has been used: isometric matching of the 2-surface $S$.
This can be expressed in terms of quasi-spherical foliation adapted coordinates $t, r, \theta, \varphi$ as
\begin{equation}
g_{AB} \doteq \bar g_{AB} = \bar g_{ij} y^i_A y^j_B = -y^0_A y^0_B + \delta_{ij} y^i_A y^j_B\,, \label{2Diso}
\end{equation}
where $S$ is given by constant values of $t, r$, and $A, B$ range over $\theta, \varphi$.  We use $\doteq$ to indicate a relation which holds only on the 2-surface $S$.
Eq.~(\ref{2Diso}) is 3 conditions on the 4 functions $y^i$.  One can regard $y^0$ as the free choice.
From a classic closed 2-surface into $\mathbb R^3$ embedding theorem---as long as $S$ and $y^0(x^\mu)$ are such that on $S$
\begin{equation}
g_{AB}' := g_{AB} + y^0_A y^0_B
\end{equation}
is convex---one has a unique embedding.
Wang and Yau have discussed in detail this type of embedding of a 2-surface into Minkowski controlled by one function in their recent quasi-local work~\cite{WY09}. 


Our ``new'' proposal~\cite{Sza2000} is: complete 4-dimensional isometric matching on $S$.
This imposes 10 constraints,
\begin{equation}
g_{\mu\nu} \doteq \bar g_{\mu\nu} \doteq \bar g_{ij} y^i{}_\mu y^j{}_\nu\,, \label{4Diso}
\end{equation}
on the 16 $y^i_{~\alpha}(t_0, r_0, \theta, \varphi)$ on $S$.  On the 2-surface $S$ these 16 quantities are actually determined by 12 independent embedding functions: $y^i, y^i{}_t, y^i{}_r$ (since from $y^i$ on $S$ one can get $y^i{}_\theta, y^i{}_\varphi$),  hence there remain $2 = 12 - 10$ degrees of freedom in choosing the reference.  In detail (\ref{4Diso}) includes---in addition to the 3 components of the 2D subsector already considered in (\ref{2Diso}), which for a given $y^0$ determines $y^i_A$---7 algebraic constraints on the 8 $y^i{}_t, y^i{}_r$.  More specifically this algebraic system has effectively 2 quadratic and 5 linear constraints.  We select $y^0{}_r$ as an independent controlling function (geometrically it controls a local boost of the reference in the plane normal to the 2-surface $S$).  Requiring the existence of suitable algebraic solutions to the 7 off-surface components of (\ref{4Diso}) imposes some restrictions on the allowable controlling functions, $y^0, y^0{}_r$.  The sign choices in selecting the appropriate solution of the quadratic relations can be resolved by considering the limiting case of a flat dynamic metric.

One could as an alternative use orthonormal frames. Then the 4D isometric matching can be represented by $\vartheta^\alpha \doteq \bar\vartheta^\alpha$.  But the reference coframe has the form $\bar\vartheta^\alpha = dy^\alpha$. Thus one should Lorentz transform the coframe $\vartheta^\alpha$ to match $dy^\alpha$ on the 2-surface $S$.
This leads to an integrability condition: the 2-forms $d\vartheta^\alpha$ should vanish when restricted to the 2-surface:
\begin{equation}
d\vartheta^\alpha|_S \doteq 0,
\end{equation}
this is 4 conditions restricting the 6 parameter local Lorentz gauge freedom.  Which again shows that after 4D isometric matching there remains $2 = 6 - 4$ degrees of freedom in choosing the reference. 


There are 12 embedding variables subject to 10 4D isometric matching conditions, or equivalently, 6 local Lorentz gauge parameters subject to 4 frame embedding conditions.
To fix the remaining 2, one can regard the quasi-local value as a measure of the difference between the dynamical and the reference boundary values.  This value will be a functional of the 2 reference controlling functions $y^0, y^0{}_r$.  The critical points of this functional determine the distinguished choices for these 2 functions.

Previously we proposed~\cite{ae100} taking the optimal ``best matched'' embedding as the one which gives an extreme value to the associated invariant mass $m^2 = - p_i p_j \bar g^{ij}$.
This should determine the reference up to a Poincar\'e transformation.

This is a reasonable condition, but, unfortunately, not so practical.  The invariant mass is a sum of 4 terms, each quadratic in an integral over $S$.  Note, however, that using the Poincar\'e freedom one can get the same $m$ value in the center-of-momentum frame from $p_0$.  This leads us to our new proposal: take the preferred reference as one that gives a critical value to the quasi-local \emph{energy} given by (\ref{E(N,S)}) and (\ref{B2}) with $N^k = \alpha^k = \delta^k_0$.  We expected this much simpler optimization to give the same reference geometry as that obtained from using $m^2$.

Based on some physical and practical computational arguments it seems reasonable to expect a unique solution in general.  In a numerical calculation in principle one could just calculate the energy values given by (\ref{E(N,S)}) and (\ref{B2}) with $N^k = \alpha^k = \delta^k_0$ for a great many choices of $y^0, y^0{}_r$ subject to the 4D isometric matching constraint (\ref{4Diso}) and the integrability condition $y^i{}_A = \partial_A y^i$ and then note the energy critical points.

Analytically the procedure is complicated by the lack of an explicit formula for the general solution of the 2D isometric embedding.  For certain cases with special symmetry this is not an obstacle.
This ``best matching'' procedure already gave reasonable quasi-local energy results for spherically symmetric systems~\cite{WCLN},
and we now have sensible results for certain axisymmetric systems including the Kerr metric~\cite{cjp13}.

Our objective was just to find a good way to select the reference for the Hamiltonian boundary term. Naturally this leads to values for the quasi-local quantities.  Moreover, the program has additional benefits, since the results of the construction can be applied to other unanticipated ends. These include:  a preferred coordinate frame for the Freud superpotential associated with the Einstein pseudotensor~\cite{Freud}, a preferred tetrad for the teleparallel gauge current~\cite{deAndrade:2000kr}, an optimal spinor field for the spinor Hamiltonian quasi-local boundary term~\cite{spinorham} associated with the Witten positive energy proof~\cite{Witten:1981mf}, and the ``best'' frame and spinor for the quadratic spinor Lagrangian formulation~\cite{Nester:1994zn}.

Furthermore, from a consideration of the covariant Hamiltonian
works~\cite{Nester:1991yd,Chen:1994qg,Chen:1998aw,Chang:1998wj,Chen:2000xw,Chen:2005hwa,Nester08},
we can see that our reference program (isometric embedding with critical energy value) can be used in other ways~\cite{givenN} and can be  applied in much more general settings.  The applications include
selecting the reference frame for any of the pseudotensors and the reference for the other GR boundary
terms corresponding to GR Hamiltonians with other boundary conditions. Indeed the program can be applied to all of the different boundary terms
that have been proposed for the most general metric-affine gravity theory and all its special
subcases, including the Poincar\'e gauge theory and teleparallel theory.   We plan to present detailed discussions of  these applications in future works.

\medskip

This work was supported by the National Science Council of the R.O.C. under the
grants  NSC-101-2112-M-008-006 (JMN) and NSC 99-2112-M-008-005-MY3
(CMC) and in part by the National Center of Theoretical Sciences (NCTS).



\begin{references}
\bibitem{remark}
  The 2nd term in (\ref{B}) also vanishes for 4D isometric matching on $S$, a condition we shall use  below.
\bibitem{otherrefs}
  In addition to Minkowski, other geometries could serve as a reference, preferably one with high symmetry, e.g., (Anti-)de Sitter.
\bibitem{givenN}
  One can look to the choices of the embedding variables that are  critical points of $E(N,S)$ (\ref{E(N,S)}) for a given $N$.
\end{references}
\end{document}